\documentclass[aps,pre,twocolumn,showpacs,preprintnumbers,amsmath,amssymb,nofootinbib]{revtex4}
\usepackage{graphicx}
\input epsf
\epsfclipon
\usepackage{epstopdf}
\usepackage[T1]{fontenc}

\begin{document}
\title{Mixed-order phase transition in a minimal, diffusion based spin model}
\author{Agata Fronczak and Piotr Fronczak}
\affiliation{Faculty of Physics, Warsaw University of Technology,
Koszykowa 75, PL-00-662 Warsaw, Poland}
\date{\today}

\begin{abstract}
In this paper, we exactly solve, within the grand canonical ensemble, a minimal spin model with the hybrid phase transition. We call the model  "diffusion-based" because its hamiltonian can be recovered from a simple dynamic procedure, which can be seen as an equilibrium statistical mechanics representation of a biased random walk. We outline the derivation of the phase diagram of the model, in which the triple point has the hallmarks of the hybrid transition: discontinuity in the average magnetization and algebraically diverging susceptibilities. At this point, two second-order transition curves meet in equilibrium with the first-order curve, resulting in a prototypical mixed-order behavior. 
\end{abstract} 
\maketitle


\section{Introduction}\label{SecIntro}

Spin models are widespread in science and computing. Since the beginning of the last century until the seventies these models were primarily used to understand the phenomenon of magnetism. Nowadays, they are also used in quantum information theory, theoretical computer science, and even computational sociology. Their remarkable role also lies in the fact that, despite decades of studies, the models are still a rewarding subject of research which continuously, for many years, provides new interesting results in the field of phase transitions.

Recently, for example, a new one-dimensional Ising model with long range interactions has been introduced and exactly solved, in which the so-called hybrid phase transitions are observed \cite{PRLBar2014, JStatMechBar2014}. The new model is a truncated version of the prominent, inverse distance squared Ising (IDSI) model \cite{PRThouless1969}. Recall that, in the original IDSI model, ferromagnetic coupling between spins decays as $1/r^2$, where $r$ is the distance between spins. The exact solution of the original model is still not known, but many of its thermodynamic features have been precisely determined \cite{JPhysACardy1981, CMathPhysFrohlich1982, JStatPhysAizenman1988}. In particular, it was shown that there is a nontrivial transition in the model which is characterized by a discontinuous (first-order-like) jump in magnetization accompanied by exponentially diverging (second-order-like) correlation length. The coexistence of these, at first sight, opposing features in the same model has been called the "Thouless effect" \cite{PRThouless1969}. At present, the effect is also referred to as "hybrid" or "mixed-order" transition (MOT) (see e.g. \cite{PRLBaxter2012, PRECellai2013, PREBassler2015, PREJang2015, PREChmiel2015, PRLCho2016, PREFronczak2016, arxivBaxter2016}). The naming convention is relevant because it emphasizes that the celebrated dichotomy between first and second order transitions may fail in some systems, with an indication of systems  having long-range interactions.

In Refs.~\cite{PRLBar2014, JStatMechBar2014}, the authors showed that, although the exact solution of the original IDSI model is not known, one can solve its truncated version which assumes that long-range interactions only apply to spins that lie in the same domain of either up or down spins. This assumption proved to be extremely successful. The authors have not only solved the model in a nonzero magnetic field, but also showed that, there exists an abundance of phenomena similar to the Thouless effect, in the truncated model. Furthermore, they showed that, with the new assumption, the spin model becomes completely equivalent to the known Poland-Scheraga (PS) model of DNA denaturation. (To be precise, in the PS model \cite{JChPhysDNA1966, JChPhysFisher1966, PRLKafri2000}, the double-stranded DNA molecule is represented as an alternating chain of bound segments and denaturated loops, with these latter playing the role of magnetic domains.) The above mentioned equivalence consists in the fact that hamiltonians of the two models are identical and they can be written as a sum (over all domains) of local, domain-related hamiltonians, which depend logarithmically on the length of a domain.

A similar logarithmic dependence on the size of domains, leading to a phenomenon resembling the extreme Thouless effect has been recently demonstrated in yet another spin model \cite{PREFronczak2016}. The model mentioned deserves attention because of its extraordinary simplicity which refers both to the very definition of the model as well as the direct method of its complete analytical solution. We call the model "diffusion-based" because its hamiltonian can be recovered from a simple dynamic procedure, which can be seen as an equilibrium statistical mechanics representation of a biased random walk. The procedure is as follows: In the set of $N$ distinguishable and non-interacting spins, $s_i=\pm 1$, in subsequent time steps, with probability $q$, a random positive spin is changed to negative and, correspondingly, with probability $1-q$, a random negative spin is changed to positive. The relationship of the model and diffusion becomes obvious when one considers the $q-$biased one-dimensional discrete-space and -time random walker, whose position $N_+(t)$ (which corresponds to the number of positive spins) is confined by reflecting walls to the region $0\leq N_+\leq N$. The $q-$bias means that $N_+$ decreases by one with probability $q$ and increases by one with probability $1-q$, in exactly the same way as in the spin model.

In Ref.~\cite{PREFronczak2016}, the model described above was exactly solved within the canonical ensemble treatment of equilibrium statistical mechanics. It was shown that in the thermodynamic limit, for $N\rightarrow\infty$, the model has a critical point at $q_c=1/2$. Below this point, for $q<q_c$, the average magnetization per spin (being the order parameter of the model) is $\langle s\rangle_{\!ce}=+1$. (Throughout this paper, the subscript $ce$ indicates the averaging over the canonical ensemble.) Above this point, for $q>q_c$, one has $\langle s\rangle_{\!ce}=-1$, whereas for $q=q_c$ the average spin is $\langle s\rangle_{\!ce}=0$. It was also shown that the discontinuity in the order parameter is accompanied by the susceptibility, which, in finite-size systems, has an algebraic divergence: $\chi=\partial\langle s\rangle_{\!ce}/\partial q\sim |q-q_c|^{-2}/N$. This unusual behavior of the susceptibility, i.e. power-law, critical-like divergence, on the one hand, and size-dependent damping, on the other hand, makes the model an interesting subject for further research. 

\begin{figure}
 \centering \includegraphics[width=0.7\columnwidth]{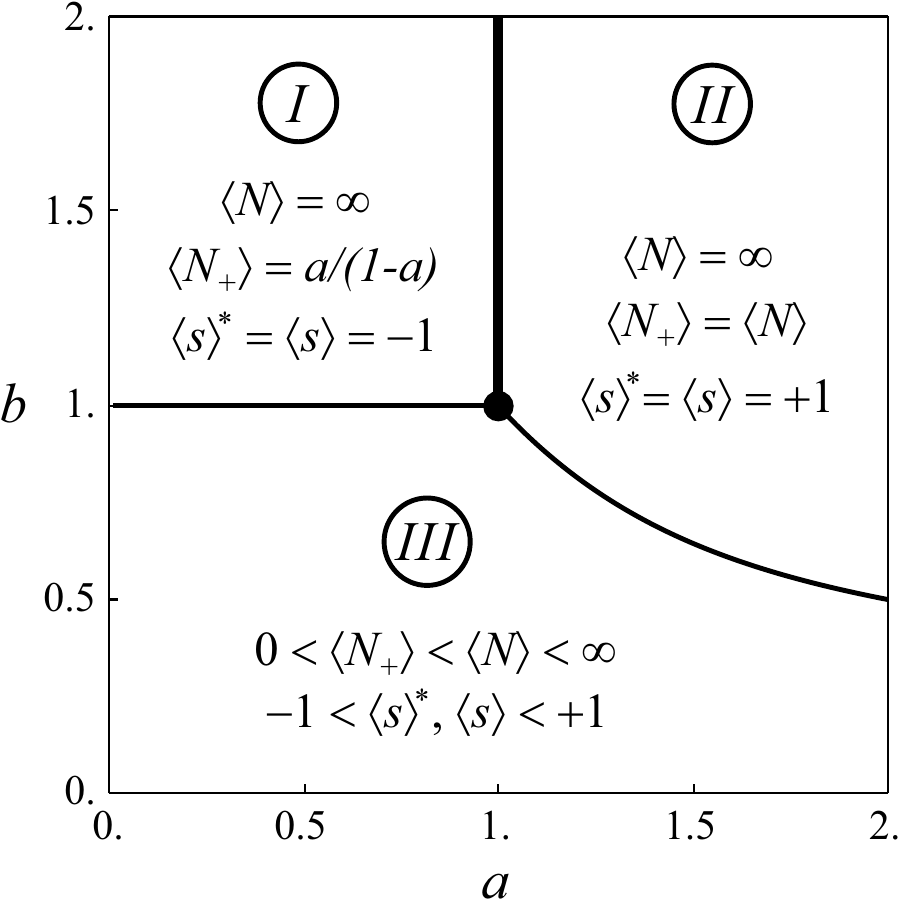}
	\caption{Phase diagram of the minimal, diffusion-based spin model. The thick line separating the totally ordered phases (I and II) is the first-order transition line. The thin curves separating the third region from the rest of the phase space are second-order curves. In the triple point, which is marked as a solid dot, mixed-order transition is observed.}
	\label{fig1}
\end{figure}

In this paper, we analyze the diffusion-based spin model in the grand canonical ensemble. This means that we investigate systems which can exchange energy and spins (particles) with a reservoir at constant (temperature-like) parameter $q$ and constant chemical potential $\mu$. We show that the phase diagram (see Fig.~\ref{fig1}) of the model is divided into three regions (phases): two of them corresponding to infinite systems with saturated magnetization, $\langle s\rangle=\pm 1$, and the third one representing systems, whose average size is finite and $\langle s\rangle$ varies continuously from $-1$ to $+1$. These three regions are separated from each other by three lines indicating phase transitions. The line between the two totally ordered phases is the line of the first-order transition. The other two lines, between the third phase and the rest of the phase space, correspond to the second-order (critical) transitions. The main result of this paper is that: The triple point at which the first-order line meets two second-order lines has a truly mixed-order nature. At this point, the hybrid transition is observed, in which the discontinuous change in the average magnetization occurs simultaneously with diverging susceptibilities, which, unlike in Ref.~\cite{PREFronczak2016}, do not depend on the system size. 

The paper is structured as follows. In Sec.~\ref{SecPrev} we briefly summarize the already known results for the minimal, diffusion based spin model that will be needed in the rest of the paper. In Sec.~\ref{SecMain}, the main result of this paper is described, which consists in detailed discussion of the phase diagram of the model in the grand canonical ensemble. Concluding remarks are given in Sec.~\ref{SecSum}.

\section{Background on the diffusion based spin model} \label{SecPrev}

In this section, we closely follow the original presentation of the model given in Ref.~\cite{PREFronczak2016}. We consider $N$ distinguishable and noninteracting spins, which can have two states, $s_i=\pm 1$. With a probability $q$ a random positive spin is chosen and flipped. Correspondingly, with a probability $1-q$ the reverse action is performed: one of negative spins is selected and turned into a positive.

The model is clearly ergodic, therefore the authors of Ref.~\cite{PREFronczak2016} concluded that there must exist its equilibrium representation in the sense of the canonical ensemble. Assuming that the probability of the system to be found in a certain spin configuration, $\Omega$, is given by the standard canonical distribution, $P(\Omega)\sim e^{-\mathcal{H}(\Omega)}$, they have noticed that the hamiltonian of the model, $\mathcal{H}(\Omega)$, can be recursively recovered from the detailed balance condition.

The hamiltonian reconstruction procedure is as follows: First, given that $\Omega$ and $\Omega'$ are two spin configurations, which differ from each other by the state of only one spin (let us say $s_i$, in such a way that $s_i(\Omega)=+1$ and $s_i(\Omega')=-1$) the transition probabilities among these configurations can be written as:
\begin{equation}\label{pO1O2}
p(\Omega\rightarrow\Omega')=\frac{q}{N_{+}(\Omega)},
\end{equation}
and
\begin{equation}\label{pO2O1}
p(\Omega'\rightarrow\Omega)=\frac{1-q}{N_{-}(\Omega')},
\end{equation}
where $N_{+}(\Omega)$ and $N_{-}(\Omega')$ stand for the number of positive and negative spins in the configurations specified, and
\begin{equation}\label{NpNm}
N_{-}(\Omega')=N-N_{+}(\Omega')=N+1-N_{+}(\Omega).
\end{equation}
Then, inserting Eqs.~(\ref{pO1O2})$-$(\ref{NpNm}) into the well-known expression for the detailed balance condition,
\begin{equation}\label{DB}
\frac{p(\Omega\rightarrow\Omega')}{p(\Omega'\rightarrow\Omega)}=\frac{P(\Omega')}{P(\Omega)} =e^{\mathcal{H}(\Omega)-\mathcal{H}(\Omega')},
\end{equation}
one gets the following recurrence relation for the hamiltonian of the model:
\begin{equation}\label{Hrec}
\mathcal{H}(\Omega)=\mathcal{H}(\Omega')+\ln\left(\frac{q}{1-q}\right)+ \ln\left(\frac{N_{-}(\Omega)+1}{N_{+}(\Omega)}\right).
\end{equation}

Eq.~(\ref{Hrec}) can be solved using standard methods (see \cite{bookConcreteMath}). It can also be solved \emph{step by step}, by imagining that we successively flip all the remaining positive spins. The flipping ends when the system reaches the ground state, $\Omega_0$, when all the spins are negative. The solution of Eq.~(\ref{Hrec}) is:
\begin{equation}\label{H0}
\mathcal{H}(\Omega)=\mathcal{H}(\Omega_0)+N_{+}(\Omega)\ln\left(\frac{q}{1-q}\right) +\ln{N\choose N_{+}(\Omega)},
\end{equation}
where $\mathcal{H}(\Omega_0)$ is a constant, which can be omitted.

With the hamiltonian given by Eq.~(\ref{H0}), the canonical partition function can be simply calculated:
\begin{eqnarray}
\label{Z00}
Z(q,N)&=&\sum_{\Omega}e^{-\mathcal{H}(\Omega)}\\
\label{Z0} &=&\sum_{\Omega} {N\choose N_+(\Omega)}^{-1}\left(\frac{1-q}{q}\right)^{N_+(\Omega)}\\
\label{Z1}&=&\sum_{N_+=0}^N a^{N_+}=\frac{1-a^{N+1}}{1-a},
\end{eqnarray}
where
\begin{equation}\label{defa}
a=\frac{1-q}{q}.
\end{equation}
Then, thermodynamic properties of the model can be obtained by successive differentiation of the partition function. In particular, one can show that the average magnetization per spin, which is given by
\begin{equation}\label{means0}
\langle s\rangle_{\!ce} =\frac{1}{N}\left\langle\sum_{i=1}^Ns_i\right\rangle_{\!\!ce} =\frac{2\langle N_+\rangle_{\!ce}}{N}-1,
\end{equation}
where
\begin{equation}\label{meanNp1}
\langle N_+\rangle_{\!ce}=a\frac{\partial\ln Z}{\partial a},
\end{equation}
shows a discontinuity in the thermodynamic limit, i.e.
\begin{equation}\label{means2}
\lim_{N\rightarrow\infty}\langle s\rangle_{\!ce}=\left\{ \begin{array}{lcl}
+1 & \mbox{for } & q<q_c \\
\;\;\;0 & \mbox{for } & q=q_c\\
-1 & \mbox{for } & q>q_c
\end{array}\right.,
\end{equation}
where
\begin{equation}\label{qc}
q_c=\frac{1}{2}.
\end{equation}

Furthermore, by differentiating $\langle s\rangle_{\!ce}$, Eq.~(\ref{means0}), with respect to $q$, one can show that, in finite-size systems, the magnetic susceptibility exhibits the mentioned intriguing behavior:
\begin{equation}\label{chi1}
\chi=\frac{\partial\langle s\rangle_{\!ce}}{\partial q} \sim\frac{1}{N}\left|q-q_c\right|^{-2}.
\end{equation}

\section{Grand canonical analysis}\label{SecMain}

\subsection{Mean-field structure of the phase diagram}

The grand canonical partition function of the model can be determined using the canonical one, see Eq.~(\ref{Z1}),
\begin{eqnarray}\label{X0}
\mathcal{Z}(q,\mu)\!&\!=\!&\!\sum_{\Omega}e^{-\mathcal{H}(\Omega)}e^{-\mu N(\Omega)}\\\label{X1}\!&\!=\!&\!\sum_{N=0}^\infty Z(q,N)b^{N}\\ \label{X2}
\!&\!=\!&\!\lim_{L\rightarrow\infty}\frac{a}{1\!-\!a}\left[\frac{1\!-\!b^{L+1}}{a(1\!-\!b)}-\frac{1\!-\!(ab)^{L+1}}{1\!-\!ab}\right],
\end{eqnarray}
where the fugacity, 
\begin{equation}\label{defb}
b=e^{-\mu},
\end{equation}
is introduced.

 \begin{figure}
 \centering \includegraphics[width=0.9\columnwidth]{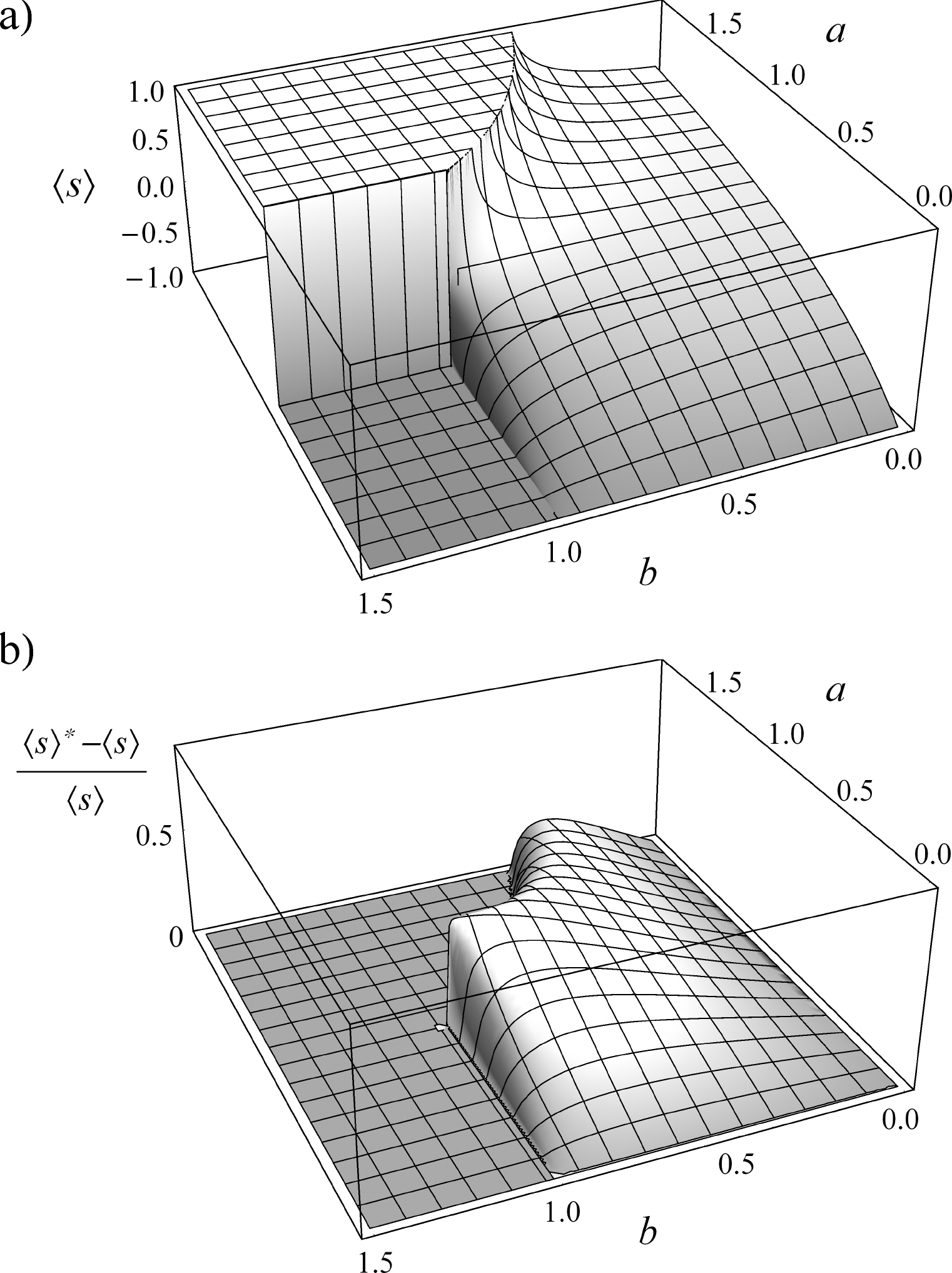}
 	\caption{a) Graph of the average magnetization per spin, $\langle s\rangle$, as given by Eq.~(\ref{means1}). b) Relative differences between the mean-field, $\langle s^*\rangle$, and the exact, $\langle s\rangle$,  order parameter, which are given by Eqs.~(\ref{meansp}) and~(\ref{means1}), respectively.}
 	\label{fig2}
 \end{figure}

Given the grand partition function, $\mathcal{Z}$, one can proceed to calculate the average system size,  $\langle N\rangle$, and the expected value of the number of positive spins,  $\langle N_+\rangle$. The averages are simply given by the derivatives of $\mathcal{Z}$:
\begin{equation}\label{meanNp2}
\langle N_+\rangle=\sum_{N=0}^{\infty}\sum_{N_+=0}^NN_+P(N_+,N)= a\frac{\partial\ln\mathcal{Z}}{\partial a},
\end{equation}
and
\begin{equation}\label{meanN2}
\langle N\rangle=\sum_{N=0}^{\infty}\sum_{N_+=0}^NNP(N_+,N)= b\frac{\partial\ln\mathcal{Z}}{\partial b},
\end{equation}
where
\begin{equation}\label{PNpN}
P(N_+,N)=\frac{a^{N_+}b^N}{\mathcal{Z}}
\end{equation}
is the probability that the system consists of $N$ spins among which $N_+$ are positive, assuming that $a$ and $b$ are fixed values. 

Inserting the grand canonical partition function, Eq.~(\ref{X2}), into Eqs.~(\ref{meanNp2}) and~(\ref{meanN2}), yields preliminary results which enable an immediate reconstruction of the phase diagram. In particular, it is easy to show that the $(a,b)$ projection of the diagram displays three distinct regions, see Fig.~\ref{fig1}. The first region (I) is located at $a<1$ and $b>1$. In this region, the expected system size is infinite, $\langle N\rangle=+\infty$, while the average number of positive spins is finite, $\langle N_+\rangle=a/(1-a)$. This makes that the mean field prediction for the average magnetization per spin, 
\begin{equation}\label{meansp}
\langle s\rangle^{\!*}=\frac{\langle\sum_is_i\rangle}{\langle N\rangle}=2\frac{\langle N_+\rangle}{\langle N\rangle}-1,
\end{equation}
gives the saturated, negative value,
\begin{equation}\label{meansp1}
\langle s\rangle^{\!*}=-1,
\end{equation}
for $a\!<\!1\;\land\;b\!>\!1$. Accordingly, in the rest of the phase space (i.e. in the  second (II) and third (III) region, as marked in Fig.~\ref{fig1}) one gets: $\langle N\rangle=\langle N_+\rangle=+\infty$ for $a\!>\!1\;\land\;b\!>\!1/a$, and $\langle N\rangle=\frac{b(1+a-2ab)}{(1-b)(1-ab)}$, $\langle N_+\rangle=\frac{ab}{1-ab}$ for $a\!<\!1/b\;\land\;b\!<\!1$. Correspondingly, the average spin is:
\begin{equation}\label{meansp2}
\langle s\rangle^{\!*}=+1,
\end{equation}
for $a\!>\!1\;\land\;b\!>\!1/a$, and
\begin{equation}\label{meansp3}
\langle s\rangle^{\!*}=\frac{a-1}{a+1-2ab},
\end{equation}
for $a\!<\!1/b\;\land\;b\!<\!1$.

The mean field description of the model, by using the average magnetization per spin as defined by Eq.~(\ref{meansp}), provides results which allow to reconstruct the phase diagram, but the results are not accurate enough to determine, what kind of phase transitions occur at the borders of the specified regions. For example, at the border of the third region, although the behavior of $\langle s\rangle^{\!*}$ resembles the behavior of the order parameter in continuous phase transitions, the magnetic susceptibilities, $\chi^*_a=\frac{\partial \langle s\rangle^{\!*}}{\partial a}|_b$ and $\chi^*_b=\frac{\partial \langle s\rangle^{\!*}}{\partial b}|_a$, 
do not behave accordingly. Both susceptibilities reach finite values at the border of this region, with one exception, namely, for $a\rightarrow 1^{-}\;\land\;b\rightarrow 1^-$ one has $\chi^*_b\sim (1-b)^{-2}$.

\begin{figure}
 \centering \includegraphics[width=\columnwidth]{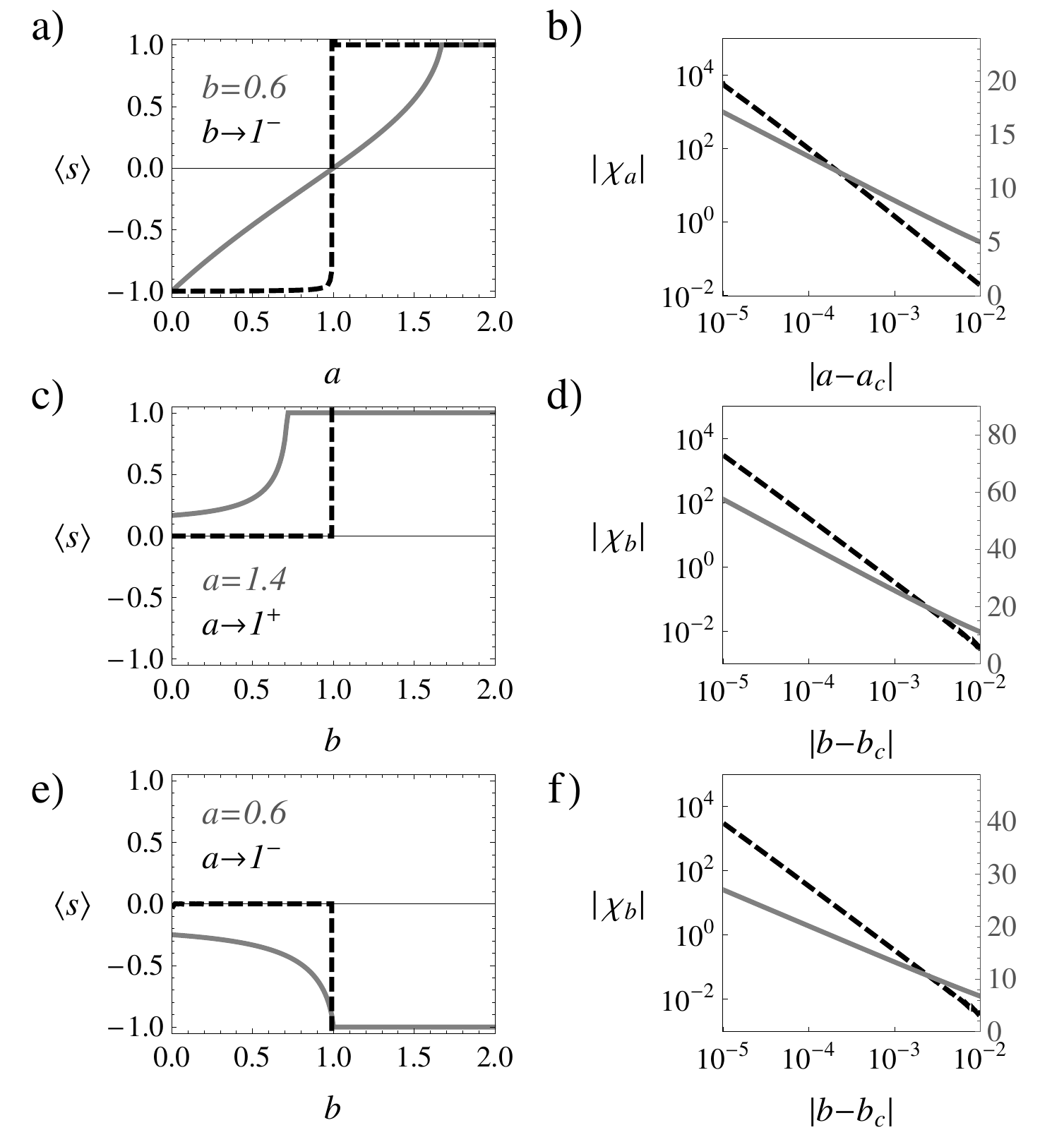}
	\caption{Average magnetization per spin, $\langle s\rangle$, and the corresponding susceptibilities, $\chi_a$ and $\chi_b$, versus different settings of the system parameters. Graphs in the same row refer to the same testing conditions, which are specified on the left plot. For example, the graphs a) and b) show variations of $\langle s\rangle$ and $\chi_a$ with $a$ for two different values of the second parameter: $b=0.60$ and $b\rightarrow 1^{-}$. In all the graphs shown in this figure, the solid curves correspond to behavior of the system in which the second-order transition is observed. Accordingly, the dashed curves correspond to the system which passes close to the triple point $a=b=1$, which is the mixed-order transition point. In the graphs showing susceptibilities, the vertical linear scales on the right side apply to solid curves only, as opposed to the logarithmic scales on their left side, which apply to dashed curves.}    
	\label{fig3}
\end{figure}

\subsection{Exact analysis of the phase diagram}

The coarse description of the model, as given by the mean field order parameter $\langle s\rangle^{\!*}$, Eq.~(\ref{meansp}), can be refined by more rigorous calculations of the average spin,
\begin{equation}\label{means1}
\langle s\rangle=\left\langle\frac{\sum_is_i}{ N}\right\rangle= 2\left\langle\!\frac{N_+}{N}\!\right\rangle-1.
\end{equation} 
To avoid the division by $N=0$ when calculating $\langle\frac{N_+}{N}\rangle$, we perform a kind of "conditional averaging" over the whole state space with the exception of the inconvenient state $N=N_+=0$. So we get:
\begin{eqnarray}\label{means2a}
\left\langle\!\frac{N_+}{N}\!\right\rangle&=&\sum_{N=1}^{\infty}\sum_{N_+=0}^N\frac{N_+}{N}P'(N_+,N)\\\label{means2b} &=&\sum_{N=1}^{\infty}\sum_{N_+=0}^N\frac{N_+}{N}\left(\frac{a^{N_+}b^N} {\sum_{R=1}^\infty Z(q,R)b^R}\right)\\\label{means2c}&=&\frac{1+A}{1+B},
\end{eqnarray}
with
\begin{equation}\label{defA}
A=\lim_{L\rightarrow\infty}\;\frac{(1-ab)}{ab(a-1)(1-(ab)^L)}\ln\left[\frac{1-ab}{1-b}\right],
\end{equation}
and
\begin{equation}\label{defB}
B=\lim_{L\rightarrow\infty}\;\frac{(1-b^L)(1-ab)}{a^2(b-1)(1-(ab)^L)}.
\end{equation}

Let us emphasize that in Eq.~(\ref{means2a}), the conditional probability distribution, $P'(N_+,N)$, is used, which is defined for $N\geq 1$, instead of the distribution $P(N_+,N)$, Eq.~(\ref{PNpN}), which is given for $N\geq 0$. Also, it is worth to point out that, the used above conditional averaging is justified in the sense that, in the vicinity of the phase boundaries, where the mean-field approach seems to fail, while the behavior of the system becomes the most interesting, one has $P(0,0)=\mathcal{Z}^{-1}\simeq 0$.

The graph of the average magnetization per spin, $\langle s\rangle$, as given by Eqs.~(\ref{means1})$-$(\ref{defB}), is shown in Fig.~\ref{fig2}a. The behavior of $\langle s\rangle$ is roughly in line with the behavior of $\langle s\rangle^{\!*}$. Therefore, the graph of $\langle s\rangle$ is consistent with the general form of the phase-diagram, which has just been obtained by the mean-field approach. There are, however, differences between $\langle s\rangle^{\!*}$ and $\langle s\rangle$ (see Fig.~\ref{fig2}b) which make that the former approach is inaccurate in more detailed analysis. As expected, the advantage of the rigorous over the mean-filed calculations arises when examining the nature of phase transitions. 

In particular, in the third region (III), where $\lim_{L\rightarrow\infty}b^L=0$ and $\lim_{L\rightarrow\infty}(ab)^L=0$, by calculating the derivatives of $\langle s\rangle$, one finds logarithmically diverging susceptibilities, when the boundary between the considered phase (III) and the totally ordered phases (I or II) is approached (see Fig.~\ref{fig3}). Thus, one has
\begin{widetext}
\begin{equation}\label{chia}
\chi_a=\left.\frac{\partial\langle s\rangle}{\partial a}\right|_{b}\!=
\frac{2(1-b)\left(ab(1-a)(2ab-a-3)+\left((1+a)(1-ab)^2+a^2(2-b-ab)\right)\ln\left[\frac{1-ab}{1-b}\right]\right)}{b(1-a)^3(ab-a-1)^2}\sim \ln|a-a_c|
\end{equation}
where 
\begin{equation}\label{ac}
a_c=\frac{1}{b}\;\;\;\;\mbox{for}\;\;\;\; b<1\;\land\;a\rightarrow a_c^{-},
\end{equation} 
and
\begin{equation}\label{chib}
\chi_b=\left.\frac{\partial\langle s\rangle}{\partial b}\right|_{a}\!=\frac{2a\left(b(1-a^2)+(2ab-a-1)\ln\left[\frac{1-ab}{1-b}\right] \right)}{b^2(1-a)^2(ab-a-1)^2}\sim\ln|b-b_c|,
\end{equation}
where
\begin{equation}\label{bc}
b_c=\frac{1}{a}\;\;\;\;\mbox{for}\;\;\;\; a>1\;\land\;b\rightarrow b_c^{-}, \mbox{\;\;\;\;and\;\;\;\;}
b_c=1\;\;\;\;\mbox{for}\;\;\;\; a<1\;\land\;b\rightarrow b_c^{-}.
\end{equation} 
\end{widetext}

Summarizing: The curves separating the third region (III) from the rest of the phase space, when they are approached from the side of smaller values of $a$ and $b$, are clearly the curves of the second-order transition. In the vicinity of these curves, excluding the triple point $a=b=1$, the order parameter, $\langle s\rangle$, changes continuously to the saturated value ($+1$ or $-1$, depending on whether $a<0$ or $a>0$), while its derivatives diverge logarithmically with the distance from these curves, cf.~Eqs.~(\ref{chia}) and~(\ref{chib}). On the other hand, the curve separating the totally ordered phases (I and II) is the first-order transition curve, which is characterized by the discontinuous jump in the order parameter and the complete lack of fluctuations. 

\subsection{Mixed-order transition}

In the phase diagram of the studied model, the triple point represents the unique conditions under which two kinds of phase transitions meet and exist together, in equilibrium. Further in this paper, we strictly show that this point has the characteristics of the hybrid (mixed-order) phase transition, i.e. discontinuity in the order parameter and diverging susceptibilities. 

Exactly at the triple point, $a=b=1$, the average magnetization per spin is 
\begin{equation}\label{mots}
\langle s\rangle=0.
\end{equation}
This result comes from elementary calculations by using Eqs.~(\ref{means1})$-$(\ref{means2b}):
\begin{equation}\label{motNpN}
\left\langle\!\frac{N_+}{N}\!\right\rangle =\sum_{N=1}^\infty\sum_{N_+=0}^N\!\! \frac{N_+}{N}\left(\frac{1^N1^{N_+}}{\sum_{R=1}^\infty\sum_{R_+=0}^R 1} \right)=\frac{1}{2},
\end{equation}
where the expression in brackets stands for the conditional probability distribution $P'(N_+,N)$. Therefore, when this point is crossed and one goes from the third region (with continuously varying magnetization) to the first or second area (in which magnetization is saturated, $\langle s\rangle=\pm 1$), then there is always a jump in magnetization. Accordingly, in the vicinity of the triple point, the both susceptibilities, $\chi_a$ and $\chi_b$, diverge as power-laws (see Fig.~\ref{fig3}):
\begin{equation}\label{motchia0}
\chi_a\sim|a-a_c|^{-2}
\end{equation}
and
\begin{equation}\label{motchib0}
\chi_b\sim |b-b_c|^{-2},
\end{equation}
with 
\begin{equation}\label{motacbc0}
a_c=b_c=1.
\end{equation}

The above scaling behavior can be deduced from Eqs.~(\ref{chia}) and~(\ref{chib}) by taking the first two terms of the following series expansion of the logarithmic function: 
\begin{equation}\label{motln}
\ln x=2\!\left[\frac{x\!-\!1}{x\!+\!1}+\frac{1}{3}\!\left(\!\frac{x\!-\!1} {x\!+\!1}\!\right)^3\!+\frac{1}{5}\!\left(\!\frac{x\!-\!1}{x\!+\!1}\!\right)^5\!+\cdots\!\right],
\end{equation}
where $x>0$. In particular, for $ab\rightarrow 1^-\;\land\;b\rightarrow 1^-$, Eq.~(\ref{chia}) can be rewritten as: 
\begin{eqnarray}\nonumber
\chi_a\!&\simeq&\! \frac{2(1-b)\left((a^2\!-\!1)+a^2(1\!-\!b)\left(\frac{1-ab}{1-b}\!+\!1\right) \ln\left[\frac{1-ab}{1-b}\right]\right)}{a(1\!-\!a)^3}\\\label{motchia1}
\!&\simeq&\!\frac{2(1\!-\!b)}{a(1\!-\!a)}+\frac{4(1\!-\!b)}{3(1\!-\!ab)^2} \;\sim\;(1-ab)^{-2}.
\end{eqnarray}
Similarly, for $a\rightarrow 1\;\land\;b\rightarrow 1^-$, Eq.~(\ref{chib}) can be rewritten as:
\begin{eqnarray}\nonumber
\chi_b\!&\simeq&\! \frac{2\left((1-a^2)-(1-b)\left(\frac{1-ab}{1-b}\!+\!1\right) \ln\left[\frac{1-ab}{1-b}\right]\right)}{(1\!-\!a)^2}\\\label{motchib1}
\!&\simeq&-2-\frac{4(1-a)}{3(2-ab-b)^2}\;\sim\;|1-b|^{-2}.
\end{eqnarray}
Note, that Eqs.~(\ref{motchia1}) and~(\ref{motchib1}) are consistent with Eqs.~(\ref{motchia0})$-$(\ref{motacbc0}).

\section{Concluding remarks}\label{SecSum}

To conclude, in this paper, we have exactly solved, within the grand canonical ensemble, the minimal, diffusion-based spin model. The model was first described in Ref.~\cite{PREFronczak2016}, where the authors have shown (using the canonical ensemble) that it exhibits an interesting behavior, which bears the hallmarks of the hybrid transition, i.e. discontinuity in the average magnetization which, in finite-size systems, is accompanied by the algebraically diverging susceptibility, see Eqs.~(\ref{means2}) and~(\ref{chi1}). The transition was tentatively named the extreme Thouless effect, but results of this paper suggest that the phenomenon should rather be recognized as a discontinuous transition, with disappearing, in the thermodynamic limit, fluctuations. Clearly, in this case, the power-law behavior of the susceptibility is not a sign of criticality. In the phase diagram, which is shown in Fig.~\ref{fig1}, a similar transition occurs on the line separating the first and second region. Accordingly, the genuine mixed-order (or hybrid) transition is encountered, when one crosses the triple point and, for example, goes from the third region to the first or second one. 

In Ref.~\cite{PREFronczak2016}, the authors have argued that the considered minimal spin model \textit{can serve as a convenient testbed for new theoretical and numerical approaches ($\dots$) dedicated to the study of mixed-order transitions}. Indeed, with the rise in the number of papers referring to the idea of hybrid transitions (see e.g. \cite{PRLBaxter2012, PRECellai2013, PREBassler2015, PREJang2015, PREChmiel2015, PRLCho2016, PREFronczak2016, arxivBaxter2016}), the problem of how to recognize the kind of transition, becomes really urgent. Trying to justify the need for such research, in Ref.~\cite{PREFronczak2016} the controversial story of explosive percolation \cite{percolScience2009, percolPRL2010a, percolPRL2010b, percolScience2011, percolNatCommun2012, percolScience2013a, percolScience2013a, percolScience2013b} has been invoked, which at first was hailed as a discontinuous transition with critical fluctuations, but then turned out to be a continuous transition. 

Another example, for which there is a chance that it could have been made a similar mistake, is the issue of the extreme Thouless effect in the minimal model of a dynamic social network, which is discussed in Refs.~\cite{PREBassler2015, EPLLiu2012, JStatMechBassler2015}. In the network model mentioned, nodes are separated into two groups representing opposing interests. Members of the first group (introverts) seek to get rid of their connections, whereas these who belong to the second group (extroverts) want to accumulate their highest possible number. It was suggested that the model exhibits the extreme Thouless effect in which the density of connections between introverts and extroverts jumps from a value which is close to zero, to a value close to unity, when the number of extroverts becomes larger than the number of introverts. Results obtained for the minimal spin model, which is as a highly simplified version of the dynamic social network, suggest that the phenomenon observed in these networks can be discontinuous phase transition with power-law fluctuations, which, however, disappear altogether in the thermodynamic limit (i.e. for large networks). 

\section{Acknowledgments}

This work has been supported by the National Science Centre of Poland (Narodowe Centrum Nauki, NCN) under grant no.~2015/18/E/ST2/00560.

\end{document}